\documentclass[12pt]{article}
\usepackage{graphicx}
\usepackage{epsfig}
\usepackage{wrapfig}
\usepackage{bm}
\usepackage{amsmath}
\usepackage{amsfonts}
\usepackage{amssymb}
\begin{document}

\title{Singlet $VA\widetilde{V}$ correlator within the instanton vacuum model}
\author{A.E. Dorokhov\\{\small Bogoliubov Laboratory of Theoretical Physics, Joint Institute for
Nuclear Research, }\\{\small 141980 Dubna, Russia}}
\maketitle
\begin{abstract}
The correlator of singlet axial-vector and vector currents in the external
electromagnetic field is studied within the instanton liquid model of QCD
vacuum. In the chiral limit we calculate the longitudinal $w_{L}^{\left(
0\right)  }$ and transversal $w_{T}^{\left(  0\right)  }$ with respect to
axial-vector index invariant amplitudes at arbitrary momentum transfer $q.$ It
is demonstrated how the anomalous longitudinal part of the correlator is
renormalized at low momenta due to the presence of the $U_{A}\left(  1\right)  $ anomaly.
\end{abstract}

\section{Introduction}

\begin{figure}[h]
\begin{center}
\includegraphics[height=4cm]{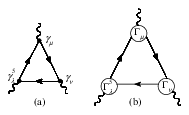}
\end{center}
\caption{{Diagrammatic representation of the triangle diagram in the local
perturbative theory (a); and in the instanton model with dressed quark lines
and full quark-current vertices (b). }}%
\label{w6}%
\end{figure}

Consideration of the axial-vector $A$ and vector $V$ current-current
correlator in the soft external electromagnetic field $\widetilde{V}$ is an
important part of the calculations of the complicated light-by-light
scattering amplitude related to the problem of accurate computation of higher
order hadronic contributions to muon anomalous magnetic moment\footnote{See,
\textit{e.g.,} \cite{APP02,CMV} and references therein.}. In this specific
kinematics when one photon $(V)$ with momentum $q_{2}\equiv q$ is virtual and
another one $(\widetilde{V})$ with momentum $q_{1}$ represents the external
electromagnetic field and can be regarded as a real photon with the
vanishingly small momentum $q_{1}$ only two invariant functions survive in
linear in small $q_{1}$ approximation. It is convenient to parameterize the
$VA\widetilde{V}$ correlator (Fig. \ref{w6}) in terms of longitudinal $w_{L}$
and transversal $w_{T}$ with respect to axial current index Lorentz invariant amplitudes%

\begin{equation}
\widetilde{T}_{\mu\nu\lambda}(q_{1},q_{2})=\frac{1}{4\pi^{2}}\left[
w_{T}\left(  q^{2}\right)  \left(  q_{2}^{2}q_{1}^{\rho}\varepsilon_{\rho
\mu\nu\lambda}-q_{2}^{\nu}q_{1}^{\rho}q_{2}^{\sigma}\varepsilon_{\rho\mu
\sigma\lambda}+q_{2}^{\lambda}q_{1}^{\rho}q_{2}^{\sigma}\varepsilon_{\rho
\mu\sigma\nu}\right)  -w_{L}\left(  q^{2}\right)  q_{2}^{\lambda}q_{1}^{\rho
}q_{2}^{\sigma}\varepsilon_{\rho\mu\sigma\nu}\right]  .\label{Tt}%
\end{equation}
Both Lorentz structures are transversal with respect to vector current,
$q_{2}^{\nu}T_{\nu\lambda}=0$. As for the axial current, the first structure
is transversal with respect to $q_{2}^{\lambda}$ while the second one is
longitudinal and thus anomalous. The appearance of the longitudinal structure
is the consequence of the Adler-Bell-Jackiw axial anomaly
\cite{Adler:1969gk,BJ}.

For the nonsinglet axial current $A^{\left(  3\right)  }$ there are no
perturbative \cite{Adler:er} and nonperturbative \cite{tHooft} corrections to
the axial anomaly and, as consequence, the invariant function $w_{L}^{\left(
3\right)  }$ remains intact when interaction with gluons is taken into
account. It was shown in \cite{VainshPLB03} that in the nonsinglet channel the
transversal structure $w_{T}^{\left(  3\right)  }$ is also free from
perturbative corrections. Nonperturbative nonrenormalization of the nonsinglet
longitudinal part follows from the 't~Hooft consistency condition
\cite{tHooft}, i.e.\ the exact quark-hadron duality realized as a
correspondence between the infrared singularity of the quark triangle and the
massless pion pole in terms of hadrons. However, for the singlet axial current
$A^{\left(  0\right)  }$ due to the gluonic $U_{A}\left(  1\right)  $ anomaly
there is no massless state even in the chiral limit. Instead, the massive
$\eta^{\prime}$ meson appears. So, one expects nonperturbative renormalization
of the singlet anomalous amplitude $w_{L}^{\left(  0\right)  }$ at momenta
below $\eta^{\prime}$ mass.

In the previous work \cite{AD05 WLT} we analyzed in the framework of the
instanton liquid model \cite{ShSh} the nonperturbative properties of the
nonsinglet triangle diagram in the kinematics specified above. We demonstrated
how the anomalous structure $w_{L}^{\left(  3\right)  }$ is saturated within
the instanton liquid model. We also calculated the transversal invariant
function $w_{T}^{\left(  3\right)  }$ at arbitrary $q$ and show that within
the instanton model at large $q^{2}$ there are no power corrections to this
structure. The nonperturbative corrections to $w_{T}^{\left(  3\right)  }$ at
large $q^{2}$ have exponentially decreasing behavior related to the short
distance properties of the instanton nonlocality in the QCD vacuum.

The present work is devoted to the study of the $U_{A}\left(  1\right)  $
anomaly effect on the singlet $VA\widetilde{V}$ correlator within the
instanton liquid model. We calculate in the chiral limit the longitudinal
$w_{L}^{\left(  0\right)  }$ and transversal $w_{T}^{\left(  0\right)  }$
invariant functions at arbitrary momentum transfer $q$ and demonstrate how the
singlet anomalous $w_{L}^{\left(  0\right)  }$ part of the correlator
$w_{L}^{\left(  0\right)  }$ is renormalized at low momenta due to presence of
the $U_{A}\left(  1\right)  $ anomaly.

\section{The structure of $VA\widetilde{V}$ correlator in perturbative approach}

The amplitude for the triangle diagram can be written as a correlator of the
axial current $j_{\lambda}^{5}$ and two vector currents $j_{\nu}$ and
$\tilde{j}_{\mu}$ (Fig. \ref{w6})
\begin{equation}
\widetilde{T}_{\mu\nu\lambda}=-\int\mathrm{d}^{4}x\mathrm{d}^{4}%
y\,\mathrm{e}^{iqx-iky}\,\langle0|\,T\{j_{\nu}(x)\,\tilde{j}_{\mu
}(y)\,j_{\lambda}^{5}(0)\}|0\rangle\,,
\end{equation}
where for light $u$ and $d$ quarks one has in the local theory
\[
j_{\mu}=\bar{q}\,\gamma_{\mu}Vq,\qquad j_{\lambda}^{5}=\bar{q}\,\gamma
_{\lambda}\gamma_{5}Aq\,,
\]
the quark field $q_{f}^{i}$ has color ($i$) and flavor ($f$) indices,
$A^{\left(  0\right)  }=I,$ $A^{\left(  3\right)  }=\tau_{3}$\ are flavor
matrix of the axial current, and $V=\widetilde{V}=\frac{1}{2}\left(  \frac
{1}{3}+\tau_{3}\right)  $ are the charge matrices, with the tilted current
being for the soft momentum photon vertex.

In the local perturbative theory the one-loop result (Fig. \ref{w6}a) for the
invariant functions $w_{T}$ and $w_{L}$ for space-like momenta $q$ $\left(
q^{2}\geq0\right)  $ is
\begin{equation}
w_{L}^{\mathrm{1-loop}}=2\,w_{T}^{\mathrm{1-loop}}=2N_{c}\mathrm{Tr}\left(
AV\widetilde{V}\right)  \int_{0}^{1}\frac{\mathrm{d}\alpha\,\alpha(1-\alpha
)}{\alpha(1-\alpha)q^{2}+m_{f}^{2}}\,, \label{wlt}%
\end{equation}
where $N_{c}$ is the color number, and for light quark masses one takes
$m_{f}\equiv m_{u}\approx m_{d}$. In the chiral limit, $m_{f}=0$, one gets the
result
\begin{equation}
w_{L}\left(  q^{2}\right)  =2w_{T}\left(  q^{2}\right)  =\frac{2N_{c}}{q^{2}%
}\mathrm{Tr}\left(  AV\widetilde{V}\right)  . \label{WLTch}%
\end{equation}

\section{The instanton effective quark model}

To study nonperturbative effects in the triangle amplitude $\widetilde{T}%
_{\mu\nu\lambda}$ at low and intermediate momenta one can use the framework of
the effective approach based on the representation of QCD\ vacuum as an
ensemble of strong vacuum fluctuations of gluon field, instantons. Spontaneous
breaking the chiral symmetry and dynamical generation of a momentum-dependent
quark mass are naturally explained within the instanton liquid model. The
instanton fluctuations characterize nonlocal properties of the QCD\ vacuum
\cite{MikhRad92,DEM97,DoLT98}. The interaction of light $u,d$ quarks in the
instanton vacuum can be described in terms of the effective 't Hooft
four-quark action with nonlocal kernel induced by quark zero modes in the
instanton field. The gauged version of the model \cite{Birse98,ADoLT00,DoBr03}
meets the symmetry properties with respect to external gauge fields, and
corresponding vertices satisfy the Ward-Takahashi identities.

In the framework of this effective model the nonsinglet $V$ and $A$
current-current correlators, the vector Adler function, the pion transition
form factor have been calculated for arbitrary current virtualities in
\cite{DoBr03,ADprdG2,AD02}. In the same model the topological susceptibility
of the QCD\ vacuum which is reduced to the singlet $A$ current-current
correlator has been considered in \cite{DoBr03,ADpepanTop}.

The spin-flavor structure of the nonlocal chirally invariant interaction of
soft quarks is given by the matrix products\footnote{The explicit calculations
below are performed in $SU_{f}(2)$ sector of the model.}
\begin{equation}
G\left(  1\otimes1+i\gamma_{5}\tau^{a}\otimes i\gamma_{5}\tau^{a}\right)
,\qquad G^{\prime}\left(  \tau^{a}\otimes\tau^{a}+i\gamma_{5}\otimes
i\gamma_{5}\right)  ,
\end{equation}
where $G$ and $G^{\prime}$ are 4-quark couplings in iso-triplet and singlet
channels. For the interaction in the form of 't Hooft determinant one has the
relation $G^{\prime}=-G$. In general due to repulsion in the singlet channel
the relation $G^{\prime}<G$ is required.

Within the gauged instanton model the dressed quark propagator in the chiral
limit, $S(p)$, is defined as
\begin{equation}
S^{-1}(p)=i\widehat{p}-M(p^{2}),\label{QuarkProp}%
\end{equation}
with the momentum-dependent quark mass
\begin{equation}
M(p^{2})=M_{q}f^{2}(p^{2})\label{Mp}%
\end{equation}
found as the solution of the gap equation
\begin{equation}
M(p^{2})=4G_{P}N_{f}N_{c}f^{2}(p^{2})\int\frac{d^{4}k}{\left(  2\pi\right)
^{4}}f^{2}(k^{2})\frac{M(k^{2})}{D(k^{2})},\label{SDEq}%
\end{equation}
where we denote
\[
D\left(  k^{2}\right)  =k^{2}+M^{2}(k).
\]
The constant $M_{q}\equiv M(0)$ in (\ref{Mp}) is determined dynamically from
Eq.~(\ref{SDEq}) and the function $f(p)$ defines the nonlocal kernel of the
four-quark interaction.  Within the instanton model $f(p)$ describing the
momentum distribution of quarks in the nonperturbative QCD vacuum is expressed
through the quark zero mode function. It is implied in \cite{AD05
WLT,DEM97,DoLT98} that the quark zero mode in the instanton field is taken in
the axial gauge when the gauge dependent dynamical quark mass is defined. In
particular it means that $f(p)$ for large arguments decreases like some
exponential in $p^{2}.$ To make numerics simpler we shell use the Gaussian
form
\begin{equation}
f(p)=\exp\left(  -p^{2}/\Lambda^{2}\right)  ,\label{MassDyna}%
\end{equation}
where the parameter $\Lambda$ characterizes the size of nonlocal fluctuations
in the QCD vacuum and it is proportional to the inverse average size of an instanton.

\begin{figure}[h]
\begin{center}
\includegraphics[width=10cm]{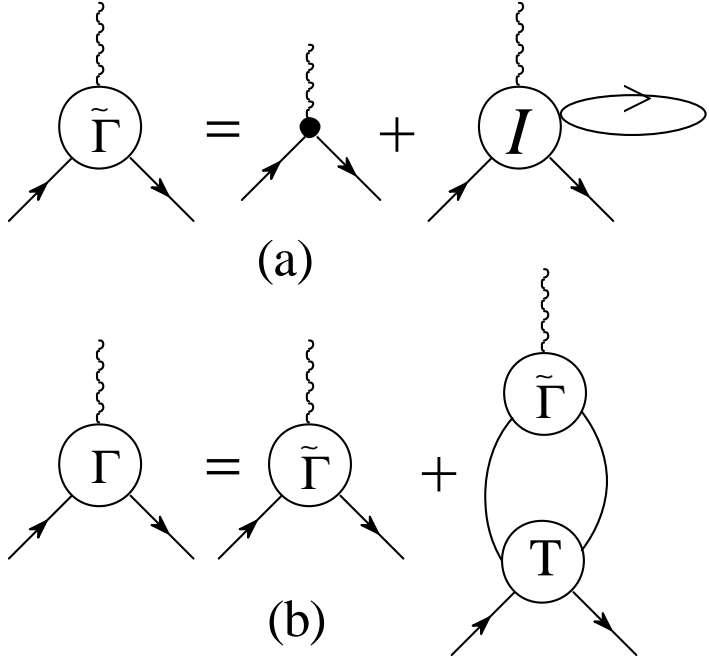}
\end{center}
\caption{{Diagrammatic representation of the bare (a) and full (b)
quark-current vertices. }}%
\label{w5}%
\end{figure}

The vector vertex following from the instanton model is \cite{ADoLT00,DoBr03}
(Fig. \ref{w5}a)
\begin{equation}
\Gamma_{\mu}(k,k^{\prime})=\gamma_{\mu}+(k+k^{\prime})_{\mu}M^{(1)}%
(k,k^{\prime}), \label{GV}%
\end{equation}
where $M^{(1)}(k,k^{\prime})$ is the finite-difference derivative of the
dynamical quark mass, $q$ is the momentum corresponding to the current, and
$k$ $(k^{\prime})$ is the incoming (outgoing) momentum of the quark,
$k^{\prime}=k+q$. The finite-difference derivative of an arbitrary function
$F$ is defined as
\begin{equation}
F^{(1)}(k,k^{\prime})=\frac{F(k^{\prime})-F(k)}{k^{\prime2}-k^{2}}.
\label{FDD}%
\end{equation}

The nonlocal part of the vertex (\ref{GV}) necessarily appears in
order to fit the vector Ward-Takahashi identity.

Within the chiral quark model \cite{ADoLT00} based on the non-local structure
of instanton vacuum \cite{DEM97} the full singlet axial-vector vertex
including local and nonlocal pieces is given by \cite{DoBr03}
\begin{align}
\Gamma_{\mu}^{5\left(  0\right)  }(k,q,k^{\prime}=k+q) &  =\gamma_{\mu}%
\gamma_{5}+\gamma_{5}(k+k^{\prime})_{\mu}M_{q}\frac{\left(  f\left(
k^{\prime}\right)  -f\left(  k\right)  \right)  ^{2}}{k^{\prime2}-k^{2}%
}-\label{G50}\\
&  +\gamma_{5}\frac{q_{\mu}}{q^{2}}2M_{q}f\left(  k^{\prime}\right)  f\left(
k\right)  \frac{G^{\prime}}{G}\frac{1-GJ_{PP}(q^{2})}{1-G^{\prime}J_{PP}%
(q^{2})}.\nonumber
\end{align}
where
\begin{equation}
J_{PP}(q^{2})=\frac{8N_{c}}{M_{q}^{2}}\int\frac{d^{4}k}{\left(  2\pi\right)
^{4}}\frac{M_{+}M_{-}\left(  k_{+}\cdot k_{-}+M_{+}M_{-}\right)  }{D_{+}D_{-}%
}.\label{Ppp}%
\end{equation}
Here and below we use the notations
\begin{align*}
k_{+} &  =k,\qquad k_{-}=k-q,\qquad k_{\perp}^{2}=k_{+}k_{-}-\frac{\left(
k_{+}q\right)  \left(  k_{-}q\right)  }{q^{2}},\\
M_{\pm} &  =M(k_{\pm}^{2}),\ \ \ \ D_{\pm}=k_{\pm}^{2}+M_{\pm}^{2},\qquad
f_{\pm}=f(k_{\pm}^{2}).\ \ \ \
\end{align*}

The vertex (\ref{G50}) takes into account the quark-antiquark rescattering in
the singlet axial channel (Fig. \ref{w5}b). For completeness we also present
the vertex corresponding to the conserved iso-triplet axial-vector current
which in the chiral limit is given by \cite{ADoLT00}
\begin{align}
\Gamma_{\mu}^{5\left(  3\right)  }(k,k^{\prime}) &  =\tau_{3}\left[
\gamma_{\mu}\gamma_{5}+2\gamma_{5}\frac{q_{\mu}}{q^{2}}M_{q}f(k)f(k^{\prime
})\right.  \label{GAtot}\\
&  \left.  +\gamma_{5}(k+k^{\prime})_{\mu}M_{q}\frac{\left(  f(k^{\prime
})-f\left(  k\right)  \right)  ^{2}}{k^{\prime2}-k^{2}}\right]  .\nonumber
\end{align}

The isotriplet axial-vector vertex has a massless pole, $q^{2}=0$, that
follows from the spontaneous breaking of the chiral symmetry in the limit of
massless $u$ and $d$ quarks. Evidently, this pole corresponds to the massless
Goldstone pion.

The singlet current (\ref{G50}) does not contain massless pole due to presence
of the $U_{A}\left(  1\right)  $ anomaly. Indeed, as $q^{2}\rightarrow0$ there
is compensation between denominator and numerator in (\ref{G50})
\begin{equation}
\frac{1-GJ_{PP}(q^{2})}{-q^{2}}=G\frac{f_{\pi}^{2}}{M_{q}^{2}}\qquad
\mathrm{as}\quad q^{2}\rightarrow0,\label{NoGold}%
\end{equation}
where $f_{\pi}$ is the pion weak decay constant. In cancellation of the
massless pole the gap equation is used. Instead, the singlet current develops
a pole at the $\eta^{\prime}-$ meson mass\footnote{See previous footnote. Also
we neglect the effect of the axial-pseudoscalar mixing with the longitudinal
component of the flavor singlet $f_{1}$ meson.}
\begin{equation}
1-G^{\prime}J_{PP}(q^{2}=-m_{\eta^{\prime}}^{2})=0,\label{Eta1}%
\end{equation}
thus solving the $U_{A}(1)$ problem. Let us also remind that in the instanton
chiral quark model the connection between the gluon and effective quark
degrees of freedom is fixed by the gap equation.

\section{Singlet $VA\widetilde{V}$ correlator}

In the effective instanton-like model the nondiagonal correlator of vector
current and singlet axial-vector current in the external electromagnetic field
($VA\widetilde{V}$) is given by (Fig. \ref{w6}b)
\begin{align}
\widetilde{T}_{\mu\nu\lambda}(q_{1},q_{2})  &  =-2N_{c}\mathrm{Tr}\left(
AV\widetilde{V}\right)  \int\frac{d^{4}k}{\left(  2\pi\right)  ^{4}}%
\cdot\label{Tncqm}\\
\cdot &  \mathrm{Tr}\left[  \Gamma_{\mu}\left(  k+q_{1},k\right)  S\left(
k+q_{1}\right)  \Gamma_{\lambda}^{5\left(  0\right)  }\left(  k+q_{1}%
,k-q_{2}\right)  S\left(  k-q_{2}\right)  \Gamma_{\nu}\left(  k,k-q_{2}%
\right)  S\left(  k\right)  \right]  ,\nonumber
\end{align}
where the quark propagator, the vector and the axial-vector vertices are
defined by (\ref{QuarkProp}), (\ref{GV}) and (\ref{G50}), respectively. The
structure of the vector vertex (\ref{GV}) guarantees that the amplitude is
transversal with respect to vector indices
\[
\widetilde{T}_{\mu\nu\lambda}(q_{1},q_{2})q_{1}^{\mu}=\widetilde{T}_{\mu
\nu\lambda}(q_{1},q_{2})q_{2}^{\nu}=0
\]
and the Lorentz structure of the amplitude is given by (\ref{Tt}).

In \cite{AD05 WLT} we found for the nonsinglet axial current $\Gamma_{\lambda
}^{5\left(  3\right)  }$ the expressions for the longitudinal amplitude
\begin{equation}
w_{L}^{\left(  3\right)  }(q^{2})=\frac{2N_{c}}{3}\frac{1}{q^{2}%
},\label{A4Tot}%
\end{equation}
and for the combination of invariant functions which shows up the
nonperturbative dynamics
\begin{align}
w_{LT}^{\left(  3\right)  }\left(  q^{2}\right)   &  \equiv w_{L}^{\left(
3\right)  }(q^{2})-2w_{T}^{\left(  3\right)  }(q^{2})=\frac{4N_{c}}{3q^{2}%
}\int\frac{d^{4}k}{\pi^{2}}\frac{\sqrt{M_{-}}}{D_{+}^{2}D_{-}}\left\{
\sqrt{M_{-}}\left[  M_{+}-\frac{2}{3}M_{+}^{\prime}\left(  k^{2}%
+2\frac{\left(  kq\right)  ^{2}}{q^{2}}\right)  \right]  -\right.  \nonumber\\
&  \left.  -\frac{4}{3}k_{\perp}^{2}\left[  \sqrt{M_{+}}M^{(1)}(k_{+}%
,k_{-})-2\left(  kq\right)  M_{+}^{\prime}\sqrt{M}^{\left(  1\right)  }%
(k_{+},k_{-})\right]  \right\}  ,\label{WLTf}%
\end{align}
where prime means a derivative with respect to $k^{2}$: $\ M^{\prime}%
(k^{2})=dM(k^{2})/dk^{2}$. The result (\ref{A4Tot}) which is independent of
the details of the nonlocal effective model is in agreement with the statement
about absence of nonperturbative corrections to the nonsinglet longitudinal
invariant function that follows from the 't Hooft duality arguments.

The calculations of the singlet $VA\widetilde{V}$ correlator results in the
following modification of the nonsinglet amplitudes%
\begin{align}
w_{L}^{\left(  0\right)  }(q^{2}) &  =\frac{5}{3}w_{L}^{\left(  3\right)
}\left(  q^{2}\right)  +\Delta w^{\left(  0\right)  }\left(  q^{2}\right)
,\label{WL0}\\
w_{LT}^{\left(  0\right)  }(q^{2}) &  =\frac{5}{3}w_{LT}^{\left(  3\right)
}\left(  q^{2}\right)  +\Delta w^{\left(  0\right)  }\left(  q^{2}\right)
,\label{WLT0}%
\end{align}
where
\begin{align}
\Delta w^{\left(  0\right)  }\left(  q^{2}\right)   &  =-\frac{5N_{c}}{9q^{2}%
}\frac{1-G^{\prime}/G}{1-G^{\prime}J_{PP}\left(  q^{2}\right)  }\int
\frac{d^{4}k}{\pi^{4}}\frac{\sqrt{M_{+}M_{-}}}{D_{+}^{2}D_{-}}\left[
M_{+}-\frac{4}{3}M_{+}^{\prime}k_{\perp}^{2}-\right.  \nonumber\\
&  \left.  -M^{(1)}(k_{+},k_{-})\left(  \frac{4}{3}\frac{\left(  kq\right)
^{2}}{q^{2}}+\frac{2}{3}k^{2}-\left(  kq\right)  \right)  \right]
.\label{DW0}%
\end{align}

In \cite{AD05 WLT} it was shown that in the chiral limit the nonsinglet
transversal amplitude gets only exponentially suppressed at large momenta
corrections. The reason is that the asymptotics of the amplitude (\ref{WLTf})
is proportional to the vacuum nonlocality function $f\left(  q\right)  $ that
is necessarily has exponentially decreasing asymptotics. The singlet
amplitudes differ from the nonsinglet ones by the term (\ref{DW0}) that also
has exponentially suppressed large $q^{2}$ asymptotics. Thus, within the
instanton model the singlet longitudinal and transversal parts have only
exponentially suppressed at large $q^{2}$ corrections and all allowed by
operator product expansion power corrections are canceled each other.

\begin{figure}[h]
\hspace*{1cm} \begin{minipage}{7cm}
\vspace*{0.5cm} \epsfxsize=6cm \epsfysize=5cm \centerline{\epsfbox{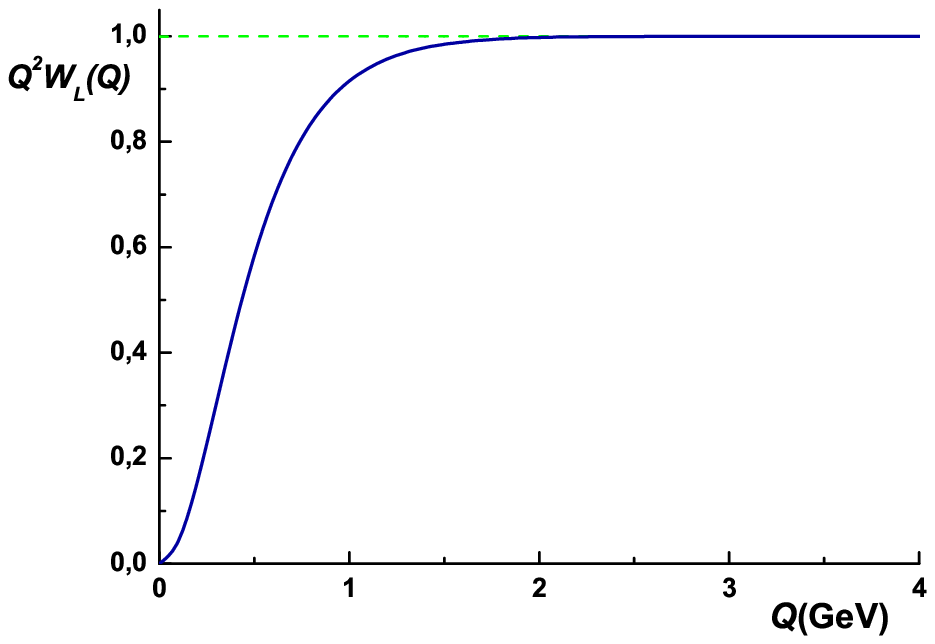}}
\caption[dummy0]{ Normalized $w_L$
invariant function in the singlet case (solid line)
and nonsinglet case (dashed line).
\label{WLfig} }
\end{minipage}\hspace*{0.5cm} \begin{minipage}{7cm}
\vspace*{0.5cm} \epsfxsize=6cm \epsfysize=5cm \centerline{\epsfbox
{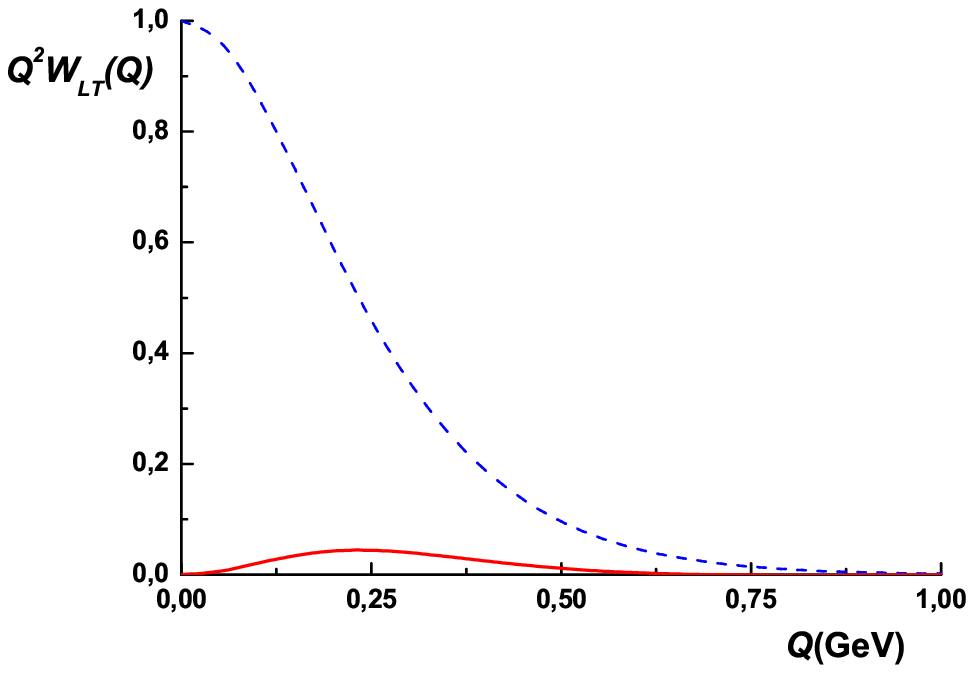}}
\caption[dummy0]{ Normalized $w_{LT}$
invariant function  versus
$Q$ predicted by the instanton model in the singlet case (solid line)
and isotriplet case (dashed line).
\label{WLTfig} }
\end{minipage}
\end{figure}

Fig. 3 illustrates how the singlet longitudinal amplitude $w_{L}^{\left(
0\right)  }$ is renormalized at low momenta by the presence of the
$U_{A}\left(  1\right)  $ anomaly. The behavior of $w_{LT}^{\left(  0\right)
}(q^{2})$ is presented in Fig. 4. In both figures the corresponding results
for the nonsinglet case are also shown. The values of the model parameters
used in calculations are fixed earlier in \cite{AD05 WLT,ADprdG2} as
\begin{equation}
M_{q}=0.24~\mathrm{GeV,}\qquad\Lambda_{P}=1.11~\mathrm{GeV,\quad}%
G_{P}=27.4~\mathrm{GeV}^{-2}.\label{G's}%
\end{equation}
The coupling $G^{\prime}$ is fixed by fitting the meson spectrum.
Approximately one has $G^{\prime}\approx0.1~G$ \cite{Birse98}. We also find
numerical values of the invariant amplitudes at zero virtuality%
\begin{equation}
w_{L}^{\left(  0\right)  }(q^{2}=0)=4.4~\mathrm{GeV}^{-2},\qquad
w_{LT}^{\left(  0\right)  }(q^{2}=0)=0.6~\mathrm{GeV}^{-2}.
\end{equation}
Precise form and even sign of $w_{LT}^{\left(  0\right)  }(q^{2})$ strongly
depend on the ratio of couplings $G^{\prime}/G$ and has to be defined in the
calculations with more realistic choice of model parameters.

\section{Conclusions}

In the framework of the instanton liquid model we have calculated for
arbitrary momenta transfer the nondiagonal correlator of the singlet
axial-vector and vector currents in the background of a soft vector field. For
this specific kinematics we find that in the chiral limit the large momenta
power corrections are absent for both longitudinal $w_{L}$ as well transversal
$w_{T}$ invariant amplitudes. These amplitudes have very similar behavior and
are corrected only by exponentially small terms which reflect the nonlocal
structure of QCD vacuum.

Within the instanton model the renormalization of the singlet longitudinal
$w_{L}^{\left(  0\right)  }$ amplitude occurring at low momenta due to the
$U_{A}\left(  1\right)  $ anomaly is demonstrated explicitly. In the
nonsinglet case the behavior of $w_{L}$ and $w_{T}$ at low momenta is very
different due to the contribution of the massless pion state. At the same time
in the singlet case there is no massless state and the deflection of $w_{L}$
from $2w_{T}$ amplitudes is rather small.

The author is grateful to A. P. Bakulev, N. I. Kochelev, P. Kroll, S. V.
Mikhailov, A. A. Pivovarov, O. V. Teryaev for helpful discussions on the
subject of the present work. The author also thanks for partial support from
the Russian Foundation for Basic Research projects nos. 03-02-17291,
04-02-16445 and the Heisenberg--Landau program.

\end{document}